%% file: main.tex
\documentclass[conference,10pt]{IEEEtran}

\usepackage{amsmath}    
\usepackage{graphicx}   
\usepackage{epstopdf}
\usepackage{verbatim}   
\usepackage{color}      
\usepackage{subfigure}  
\usepackage{algorithm}
\usepackage{array}
\usepackage{algorithmic}
\usepackage{nomencl}
\usepackage{graphicx}
\usepackage{amssymb}
\usepackage{enumerate}
\usepackage{multirow}
\usepackage{listings}
\usepackage{url}
\usepackage[hidelinks]{hyperref}
\usepackage{rotating}
\usepackage{wasysym}
\usepackage{footmisc}

\newtheorem{example}{Example}

\newtheorem{definition}{Definition}

\def\etal{\textit{et al.}}

\def\toolname{Y2U}
\def\yakindu{Yakindu}
\def\uppaal{UPPAAL}

\def\map{\mathcal{M}}
\def\medicalGuideline{\mathcal{G}}
\def\resMap{\mathcal{M}_{\res}}
\def\avaMap{\mathcal{M}_{a}}
\def\resSet{R}
\def\res{r}
\def\state{S}
\def\transition{T}
\newcommand{\var}[1]{V_{#1}} 
\def\guard{G}
\def\timeinterval{T}
\def\starttime{t_s}
\def\endtime{t_e}
\def\schedule{S}
\def\avaModel{\mathcal{A}}
\def\resAND{\otimes}
\def\resOR{\oplus}
\newcommand{\resSEQ}[1]{\odot_{#1}} 
\def\resDemand{d}

\def\procedure{p}
\def\procedureSet{P}
\def\architectureName{SMJV}
\def\architectureFullName{\textit{separately model and jointly verify}}

\IEEEoverridecommandlockouts

\begin{document}

\title{Model and Integrate Medical Resource Available Times and Relationships in Verifiably Correct Executable Medical Best Practice Guideline Models\\ (Extended Version)}

\author{
	\IEEEauthorblockN{Chunhui Guo, Zhicheng Fu, Zhenyu Zhang, Shangping Ren\thanks{This work is supported in part by NSF CNS 1545008 and NSF CNS 1545002.}}
	\IEEEauthorblockA{Department of Computer Science\\
		Illinois Institute of Technology\\
		Chicago, IL 60616, USA\\
		\{cguo13, zfu11, zzhang111\}@hawk.iit.edu, ren@iit.edu}
	\and
	\IEEEauthorblockN{Lui Sha}
	\IEEEauthorblockA{Department of Computer Science\\
		University of Illinois at Urbana-Champaign\\
		Urbana, IL 61801, USA\\
		lrs@illinois.edu}
}

\maketitle

\begin{abstract}
\input{abs}
\end{abstract}

\section{Introduction}
\label{sec:intro}
\input{intro}

\section{Related Work}
\label{sec:related}
\input{related}

\section{Verifiably Correct Executable Medical Best Practice Guidelines}
\label{sec:statechart}
\input{statechart}

\section{Annotate Medical Resource Demands}
\label{sec:annotation}
\input{annotation}

\section{Model Medical Resource Demands and Their Relationships}
\label{sec:resource}
\input{resource}

\section{Model Medical Resource Available Times}
\label{sec:availability}
\input{availability}

\section{Integrate Medical Resource Models with Medical Guideline Statecharts}
\label{sec:integration}
\input{integration}

\section{Simplified Stroke Case Study}
\label{sec:exp}
\input{exp}

\section{Conclusion}
\label{sec:conclusion}
\input{conclusion}


\bibliographystyle{abbrv}
\bibliography{ref}

\end{document}

%% file: abs.tex
Improving patient care safety is an ultimate objective for medical cyber-physical systems. A recent study shows that the patients' death rate is significantly reduced by computerizing medical best practice guidelines~\cite{Mckinley2011computer}. Recent data also show that some morbidity and mortality in emergency care are directly caused by delayed or interrupted treatment due to lack of medical resources~\cite{delayreport}.
However, medical guidelines usually do not provide guidance on medical resource demands and
how to manage potential unexpected delays in resource availability.
If medical resources are temporarily unavailable, safety properties in existing executable medical guideline models may fail which may cause increased risk to patients under care.

The paper presents a \textit{\architectureFullName} (\architectureName) architecture to separately model medical resource available times and relationships and jointly verify safety properties of existing medical best practice guideline models with resource models being integrated in. 
The \architectureName\ architecture allows medical staff to effectively manage
medical resource demands and unexpected resource availability delays during emergency care.
The separated modeling approach also allows different domain professionals to make independent model modifications,
facilitates the management of frequent resource availability changes,
and enables resource statechart reuse in multiple medical guideline models. A simplified stroke scenario is used as a case study to investigate the effectiveness and validity of the \architectureName\ architecture. The case study indicates that the \architectureName\ architecture
is able to identify unsafe properties caused by unexpected resource delays.

%% file: intro.tex
Patient care safety is time critical, especially for emergency care. Improving patient care safety needs to consider dynamic patient conditions, proper treatment procedures, treatment timing constraints, medical resource demands, and unexpected resource delays. Due to the complicated interdependencies among patient conditions, treatments and their corresponding timing constraints, and needed medical resources, it is difficult for medical professionals to consider and process all the dynamic information and make quick and accurate decisions. Hence, a computer system is needed to assist medical professionals in validating and verifying the medical guideline models with all the information being taken into considerations.

\begin{figure}[ht]
	\centering
	\includegraphics[width = 0.3\textwidth]{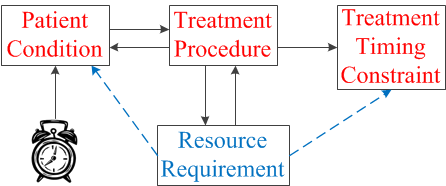}
	\caption{Interdependencies}
	\label{fig:relation}
\end{figure}

Medical best practice guidelines are intended to optimize patient care, that are informed by a systematic review of evidence and an assessment of the benefits and harms of alternative care options. For an acute illness, there is often a \textit{golden time} window during which prompt  and proper medical treatments have the highest likelihood for optimal patient care outcomes~\cite{goldenTime}. As patient condition can change rapidly, delayed or interrupted medical treatment procedures will often result in increased morbidity and mortality. Doctors and researchers at the Ottawa Hospital Research Institute studied 15,160 patients waiting for emergency medical procedures and found that of those 2820 patients who experienced delays, 138 died due to lack of timely medical care, such as delay due to lack of operating room resources and medical devices~\cite{delayreport}. We use a stoke patient as an example to show the complex interdependencies among patient condition, proper treatments and their corresponding \textit{golden time}, and needed medical resources depicted in Fig.~\ref{fig:relation}.

\textbf{Stroke Scenario:}\textit{
An ischemic stroke occurs when a clot or a mass blocks a blood vessel cutting off blood flow to a part of the brain and results in a corresponding loss of neurological function~\cite{IschemicStroke}. The intravenous (IV) tissue plasminogen activator (tPA) injection is a standard treatment for ischemic stroke patients and it is most effective during the initial 3-hour window from the onset of stroke symptoms~\cite{tpa-gold-standard}. The treatment window can be extended from 3 to 4.5 hours for certain patients, but the risks are increased~\cite{StrokeGuideline}. Some patients can be treated by dripping tPA directly on the clot through a micro-catheter within 6 hours from the onset of stroke symptoms~\cite{Prince2013strokeIA}. However, the micro-catheter tPA treatment requires specialists to control tPA dose, special equipments to put the micro-catheter into blood vessels, and technicians to operate the special equipment.}

\textit{In addition, in order to use the tPA treatment, we must ensure that (1) CT scan does not show hemorrhage,  and (2) the patient's blood pressure is under control. To derive the conclusion that the patient does not have hemorrhage, we would need medical resources including a CT machine, a CT technician, and a radiologist. If a patient's blood pressure is not within the range for tPA administration, a specialist is required to control blood pressure.} 

\setlength{\intextsep}{5pt}
\setlength{\columnsep}{0pt}

As the stroke scenario indicates, for many acute emergency care, taking proper treatment
actions is a challenging task.
Recent study conducted by Houston Methodist Hospital Research Institute shows that computerized executable medical best practice guidelines can significantly
help physicians to make treatment decisions, improve patient care safety, and reduce patients' death rate (for sepsis patients, the death rate drops from 31\% to 14\% by computerizing a sepsis best practice protocol)~\cite{Mckinley2011computer}.
For this reason, many computer executable medical best practice guideline models are developed over the past decade to assist medical professionals in medical practices~\cite{Balser2002Asbru,patel1998representing,Tu2001EON,fox1998disseminating,Terenziani2004GLARE}.

As medical guideline handbooks often do not provide instructions on medical
resource demands and how to manage unexpected delays in resource availability,
their computerized models based on the medical handbooks often also lack these information.
In elective medical procedures, patient's treatments are scheduled according to resource availability. However, it is not possible to schedule patient treatments in emergency care. If medical resources are temporarily unavailable for emergency care, certain transitions in computerized medical guideline models may be blocked, which can falsify validated/verified safety properties. In such cases, medical staff have to improvise treatment decisions, which increases patient's risk.

Take a stroke patient as an example, assume a stroke patient's onset time is 0 and a physician orders a CT scan for the patient at time 20 (minutes). If the CT machine is available, the tPA administration can be completed within the 3-hour window. However, if the CT machine or a radiologist is not available for 200 minutes,  the tPA administration will not be completed within 3-hour time window. Hence, taking considerations of required medical resources' available time in validating and verifying executable medical guideline models is
essential for patient care safety.

The resource available time is often given in a timetable~\cite{Perez2011HSE}. Medical resources by themselves often do not have any relationships, but when they are associated with medical treatment procedures, the procedures may require certain relationships among the resources.  For the CT scan scenario in the stroke example, obtaining CT results contains two main sequential steps, i.e., taking CT image followed by image diagnosis. The imaging step requires both a CT machine and a CT technician at the same time; while a radiologist is needed to diagnose images after the image is obtained. Hence, the relationships between machines and technicians and between machines/technicians and radiologists are concurrent and sequential, respectively.

It is worth pointing out that the study of temporarily unavailable medical resources
is a challenging medical cyber-physical problem.
The interruption to a medical procedure is different from interruptions handling in computer
science.
The medical resource available time and their relationship issue has to be addressed in a systematic way and in conjunction with consideration of dynamic patient conditions, medical treatment procedures, and treatment timing constraints. It involves domain knowledge from three areas: (1) medical resource available time which is usually managed by medical schedulers, (2) illness and treatments which is usually managed by medical doctors; and (3) safety verification  of computerized models done by computer scientists. 

\begin{figure}[ht]
	\centering
	\includegraphics[width = 0.49\textwidth]{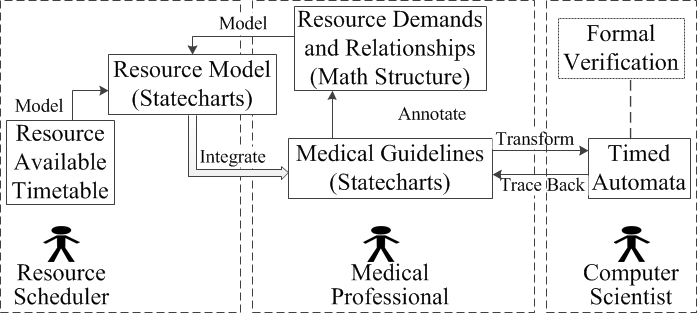}
	\caption{\architectureName\ Architecture}
	\label{fig:architecture}
\end{figure}

The paper presents a \textit{\architectureFullName} (\architectureName)   architecture to separately model medical resource available times and relationships and jointly verify safety properties of existing medical best practice guideline models with resource models being integrated in. This architecture is built on top of an existing framework which transforms validated medical best practice guideline statechart models  to \uppaal\ time automata with \toolname\ tool~\cite{Guo2016ICCPS} so that the medical guidelines' safety properties can be formally verified. Fig.~\ref{fig:architecture} depicts the high level architecture of our approach. As depicted in Fig.~\ref{fig:architecture}, the \architectureName\ architecture allows different personals to focus on their own knowledge domains and make independent model modifications when needed. It requires minimal change to existing medical best practice guideline models, and can directly apply existing work/tools to address patient care safety under dynamic patient conditions by exploring different execution paths in the medical best practice guideline models with resource demands.
In addition, the separated medical resource models built with the \architectureName\ architecture can be reused by multiple medical guideline statecharts and requires minimal modification caused by resource demands and available time changes.

To our best knowledge, the \architectureName\ architecture is the first computerized formal verification architecture to automatically assist medical staff to effectively manage
medical resource demands and unexpected resource availability delays during emergency care.
Without a computerized system, engineers need to be involved in clinical care to manually modify and verify medical guideline models when
resource demands and available times change. Furthermore, the manual modifications are error prone, which increases patients' risk.

\setlength{\intextsep}{5pt}
\setlength{\columnsep}{0pt}

The rest of the paper is organized as follows.
We discuss related work in Section~\ref{sec:related}.
Section~\ref{sec:statechart} introduces a framework for building verifiably
correct executable medical guideline models.
We identify basic medical relationships, define medical resource demands, and
automatically annotate resource demands in verifiably correct executable
medical guidelines in Section~\ref{sec:annotation}.
We present the approaches to model medical resource demands and relationships
and resource available times in Section~\ref{sec:resource} and
Section~\ref{sec:availability}, respectively.
Section~\ref{sec:integration} defines the procedure for integrating medical
resource models into existing medical guideline statecharts.
A simplified stroke case study is given in Section~\ref{sec:exp} to 
illustrate the effectiveness and advantages of the presented SMJV architecture.
We draw conclusions in Section~\ref{sec:conclusion}.

%% file: related.tex
Over the past two decades, significant amount of efforts have been made in obtaining various computer-interpretable models and developing tools for the management of medical guidelines, such as GLIF~\cite{patel1998representing}, Asbru~\cite{Balser2002Asbru}, EON~\cite{Tu2001EON}, GLARE~\cite{Terenziani2004GLARE}, and PROforma~\cite{fox1998disseminating}, to name a few. To facilitate easy clinical validation, Wu \etal\ have developed a workflow adaptation approach~\cite{WuWorkflow2015} to help physicians safely adapt workflows to react to patient adverse events and a treatment validation protocol~\cite{WuTreatment2014} to enforce the correct execution sequence of performing a treatment based on precondition validations, side effects monitoring, and expected responses checking. In addition, our previous work~\cite{Guo2016ICCPS} also designed a platform to model medical guidelines with statecharts~\cite{harel1987statecharts} and automatically transform statecharts to timed automata~\cite{alur1994theory} for formal verification. These models and tools integrated clinical guidelines with the treatment flow and provided patient-specific advice when and where needed.
These tools provide a better chance of positively impacting clinician behaviors by
reducing unjustified practice variations.
However, the effects and impacts of medical treatment procedure delays due to temporarily unavailable medical resources have not been well addressed. 

Research on patients and resource scheduling is well established and growing.
Many good medical resource scheduling systems have been developed to
reduce patient waiting times and also improve the utilization of critical resources by means of tracking the availability of resources, projecting future demands for service and automating the assignment of resources to needs~\cite{ss1, Perez2011HSE, vermeulen2009adaptive, conforti2008optimization, resource_schecule}. 
For instance, the problem of scheduling patients in CT scanning department for improving resource utilization is addressed in~\cite{patrick2007improving}. A comprehensive survey of the research is provide by~\cite{cayirli2003outpatient}  and ~\cite{gupta2008appointment}.  Though scheduling software is destined for growth, integrating medical resource availabilities generated by scheduling systems with executable medical best practice guideline models to improve safety of treatments is yet to be addressed.

To study the medical resource availability issue in existing medical guideline models, Kim and Lee \etal~\cite{Kim2010TII} used timed and resource-oriented statecharts and took a direct modification approach by specifying required resource information in transition guards and as state constraints
to analyze the time and resource-constrained behavior of the system.
With similar ideas, Christov and Lori \etal~\cite{Christov2008Formally, Christov2008} used Little-JIL to model the processes in medical guidelines and represented resource as preconditions of process steps.
All these efforts have shown that medical resource availabilities is critical to the safety of executable medical best practice guideline models.
In this paper, we present an approach that separately models resource available
times and their relationships and jointly verify
the safety properties of statechart-based computerized medical best practice guideline
models with resource models being integrated in.

%% file: statechart.tex
Our previous work~\cite{Guo2016ICCPS} designed a platform to build verifiably
correct executable medical guidelines.
In particular,
we use statecharts~\cite{harel1987statecharts} to model medical guidelines and
interact with medical professionals to validate the correctness of the medical
guideline models.
The statecharts built with \yakindu\ tool~\cite{yakindu}
are then automatically transformed to \uppaal\
timed automata~\cite{behrmann2004tutorial} by the developed \toolname\
tool~\cite{Guo2016ICCPS},
so that the safety properties required by the model
can be formally verified with \uppaal.
We use the simplified stroke scenario presented in Section~\ref{sec:intro}
as an example to briefly summarize the process of building verifiably
correct executable medical guideline models.

\begin{figure}[ht]
	\centering
	\includegraphics[width = 0.49\textwidth]{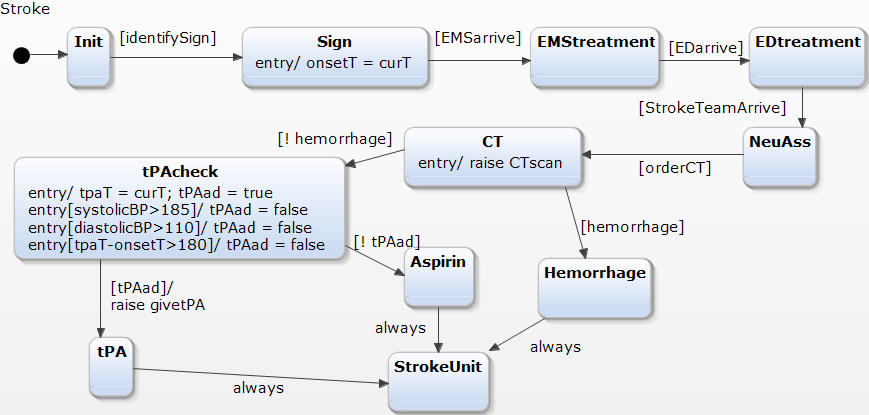}
	\caption{Simplified Stroke \yakindu\ Statechart Model}
	\label{fig:stroke}
\end{figure}

We use \yakindu\ statecharts to model the stroke treatment guideline given in~\cite{StrokeGuideline}.
For illustration and easy understanding purpose, we show a simplified stroke
statechart model in Fig.~\ref{fig:stroke}, which only focuses on the CT scan
and tPA administration procedures and omits other medical procedure details.
In the simplified statechart shown in
Fig.~\ref{fig:stroke}, we assume that upon patient arrival, treatments
to control blood pressure have been immediately performed
and the patient blood pressure is quickly brought within the safe range
if this can be done.

\setlength{\intextsep}{5pt}
\setlength{\columnsep}{0pt}

In the statechart shown in Fig.~\ref{fig:stroke}, two medical procedures $\mathtt{CTscan}$
and $\mathtt{givetPA}$ are modeled by \yakindu\ statechart as \textit{events}.
In \yakindu\ statecharts, \textit{events} can be raised
by both states and transitions.
For instance, the \textit{entry action} of state $\mathtt{CT}$ ($\mathtt{entry/ \ raise \ CTscan}$)
raises \textit{event} $\mathtt{CTscan}$ when state $\mathtt{CT}$ is entered.
The \textit{event} $\mathtt{givetPA}$ is raised by the transition from state
$\mathtt{tPAcheck}$ to state $\mathtt{tPA}$ if tPA is administrated (the value of
boolean variable $\mathtt{tPAad}$ is $\mathtt{true}$).
In the simplified stroke statechart model (Fig.~\ref{fig:stroke}),
the two time related variables $\mathtt{curT}$ and $\mathtt{onsetT}$
represent the current system time and the onset time of stroke symptoms, respectively.
We assume that the time unit in the simplified stroke statechart model is minute.
Hence, the remaining time of the 3-hour tPA treatment window can be calculated
by formula $180 - (\mathtt{curT} - \mathtt{onsetT})$.

\begin{figure}[ht]
	\centering
	\includegraphics[width = 0.49\textwidth]{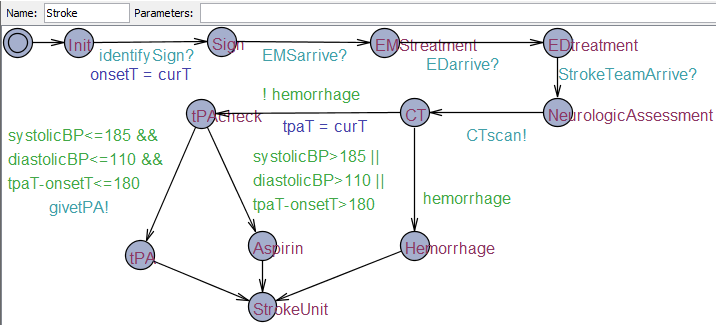}
	\caption{Simplified Stroke \uppaal\ Model}
	\label{fig:strokeU}
\end{figure}

The simplified stroke statechart model in Fig.~\ref{fig:stroke}
is transformed to \uppaal\ time automata as shown in Fig.~\ref{fig:strokeU}
with the \toolname\
tool~\cite{Guo2016ICCPS}.
There are safety properties in the simplified stroke scenario, i.e.,
\textbf{P1}: the tPA is injected only if a CT scan shows no hemorrhage
and systolic and diastolic blood pressures are smaller than or equal to
185 mm Hg and 110 mm Hg; and
\textbf{P2}: the tPA administration is completed within
3 hours from onset of symptoms.
The safety properties \textbf{P1} and \textbf{P2}
are verified in \uppaal\ by
formula~\eqref{eq:P1} and formula~\eqref{eq:P2}, respectively.

\begin{align}
	\label{eq:P1}
	\begin{split}
		A[~] \ \mathtt{Stroke.tPA} \ imply \ \mathtt{systolicBP}<=185 \ \&\& \\
		\mathtt{diastolicBP}<=110 \ \&\& \ ! \ \mathtt{hemorrhage}		
	\end{split}
\end{align}

\begin{align}
	\label{eq:P2}
	A[~] \ \mathtt{Stroke.tPAcheck} \ imply \ \mathtt{tpaT}-\mathtt{onsetT}<=180
\end{align}

\setlength{\intextsep}{5pt}
\setlength{\columnsep}{0pt}

If all the required medical resources are available,
both clinical validation
results of stroke \yakindu\ model in Fig.~\ref{fig:stroke} and
formal verification results of stroke \uppaal\ model in Fig.~\ref{fig:strokeU}
show that both properties \textbf{P1} and \textbf{P2} are satisfied.
However, if both CT machines and CT technicians are available after
200 minutes from onset of the symptoms, the stroke statechart is then blocked
at state $\mathtt{CT}$ for 200 minutes. In this scenario, the safety property
\textbf{P2} fails.

The example reveals a fact that
safety properties validated and verified in medical guideline models based on the assumption
that medical resources are available can fail because of temporarily unavailable resources.
Hence, taking into consideration of medical resource available times and relationships
in developing verifiable medical guideline models
is essential in ensuring patient care safety.
The paper addresses the issue by integrating medical resource available times and relationships
with executable medical guideline models.
In particular, in the next two sections, we first define medical resource demands with resource relationships
and automatically annotate medical resource demands in executable medical guidelines (Section~\ref{sec:annotation}),
we then model medical resource demands with statecharts (Section~\ref{sec:resource}).
In Section~\ref{sec:availability}, a mathematical structure is defined to explicitly
specify medical resource available times, then the mathematical structure is automatically
transformed to statecharts for integration with existing medical guideline statecharts.
The medical resource models are then automatically integrated
with existing medical guideline statecharts using the boolean variables built in resource statecharts
in Section~\ref{sec:integration}.
Lastly, we transform the integrated medical guideline statecharts
to timed automata~\cite{alur1994theory} by the \toolname\
tool~\cite{Guo2016ICCPS} to formally verify safety properties in the presence of temporarily
unavailable medical resources.

%% file: annotation.tex
To model medical resource demand, we need to identify which resources
are required by which medical procedures of a particular medical guideline
and resource relationships. In this section, we identify basic medical
relationships, define medical resource demand, and automatically annotate
medical resource demands in executable medical guidelines.

We use the simplified stroke statechart model shown in Fig.~\ref{fig:stroke}
as an example to illustrate medical resource demands of medical guidelines.
In the state $\mathtt{CT}$, the request for medical procedure CT scan is modeled
as $\mathtt{CTscan}$ \textit{event} in \yakindu\ statecharts. This event is raised by the \textit{entry
action} of the $\mathtt{CT}$ state. According to medical professionals, the
$\mathtt{CTscan}$ procedure requires CT machines, CT technicians, and radiologists.
The relationships between machines and technicians and between
machines/technicians and radiologists are concurrent and sequence, respectively.
Similarly, the request of giving tPA procedure is modeled as
$\mathtt{givetPA}$ event which is raised by the \textit{action}
of the \textit{transition} from state $\mathtt{tPAcheck}$ to state $\mathtt{tPA}$.
The giving tPA procedure requires tPA fluid and nurses.
The examples show that
(1) multiple medical resources required by a medical procedure have relationships;
(2) the request of medical procedures are modeled as statechart \textit{events} and can be raised
in both \textit{states} and \textit{transitions};
and
(3) treatment procedure required medical resource demands are
not represented in medical guideline models and need to be provided by medical
professionals.

\subsection{Medical Resource Relationships}
\label{subsec:relation}

To represent medical resource demands, we define three basic medical resource
relationships as follows:
\begin{enumerate}
	\item \textbf{Concurrent Resources} are resources required at the same to perform a medical
	procedure. For example, taking CT images requires both CT machines and CT technicians;
	
	\item \textbf{Alternative Resources} provide multiple resource options for a medical
	procedure, but only one option is required to perform the procedure. For example,
	either CT machines or MRI machines can provide brain images;
	
	\item \textbf{Sequence Resources} are required by a medical procedure in time sequence
	with specified time durations between them.
	For example, obtaining CT results needs a CT machine and a CT technician to take
	CT image first, and then requires a radiologist to diagnose images after 10 minutes
	(assuming it takes 10 minutes to take the CT images).
\end{enumerate}

To formally model medical resource relationships, we define three resource
relationship operators and the order of operations as follows.

\begin{definition}[Concurrent Resource Operator]
	\label{def:and}
	The concurrent resource operator $\resAND$ takes two medical resources
	$\res_1$ and $\res_2$ as arguments, i.e., $\res_1 \ \resAND \ \res_2$,
	and declares that the two medical resources $\res_1$ and $\res_2$ are concurrently required.
\end{definition}

\begin{definition}[Alternative Resource Operator]
	\label{def:or}
	The alternative resource operator $\resOR$ takes two medical resources
	$\res_1$ and $\res_2$ as arguments, i.e., $\res_1 \ \resOR \ \res_2$,
	and	declares that either medical resource $\res_1$ or $\res_2$ is required.
\end{definition}

\begin{definition}[Sequence Resource Operator]
	\label{def:seq}
	The sequence resource operator $\resSEQ{t}$ takes two medical resources
	$\res_1$ and $\res_2$ and an integer as arguments , i.e., $\res_1 \ \resSEQ{t} \ \res_2$,
	and	declares that the medical resource $\res_1$ is first required and then after
	$t$ time units resource $\res_2$ is needed, where $t \ge 0$.
\end{definition}

\begin{definition}[Operation Order of Resource Relationship Operators]
	\label{def:precedence}
	The operation order of the concurrent, alternative, and sequence resource
	relationship operators is $\resAND \ > \ \resOR \ > \ \resSEQ{t}$.
\end{definition}

It is not difficult to see that different
relationships among medical resources can be represented by the defined
three resource relationship operators. We use Example~\ref{ex:relation}
to illustrate how to represent resource relationships with the operators.

\begin{example}
\label{ex:relation}
In the simplified stroke statechart model shown in Fig.~\ref{fig:stroke}, 
a medical procedure $\mathtt{CTscan}$ is ordered in state $\mathtt{CT}$.
According to medical professionals, the $\mathtt{CTscan}$ procedure first
requires a CT machine and a CT technician to take
CT image and requires a radiologist to diagnose images after 10 minutes.
Based on Definition~\ref{def:and}, the concurrent relationship between
CT machine and CT technician is represented by $\mathtt{CT\_machine} \ \resAND \ \mathtt{CT\_technician}$.
According to sequence resource operator and operator precedence definitions,
i.e., Definition~\ref{def:seq} and Definition~\ref{def:precedence},
the sequence relationship between machine/technician and radiologist
is represented by
\begin{align}
\label{eq:relation}
\mathtt{CT\_machine} \ \resAND \ \mathtt{CT\_technician} \ \resSEQ{10} \ \mathtt{radiologist}.
\end{align}
\end{example}

\subsection{Annotate Medical Resource Demands in Executable Medical Guideline Models}
\label{subsec:annotation}

Most existing medical guideline models follow guideline handbooks and usually
do not provide instructions on medical
resource demands and how to manage unexpected delays in resource availability.
To represent medical resource demands in executable medical guideline models,
we first define medical resource demands and then automatically annotate
resource demands in executable medical guideline models.

The resource demand of a given medical procedure specifies the procedure required
medical resources and their relationships.
We give the formal definition of medical resource demands and an example
of the CT scan scenario as follows.

\begin{definition}[Resource Demand]
	\label{def:demand}
	The resource demand $\resDemand$ of a given medical procedure $\procedure$
	is defined as an expression of medical resources $\resSet$ and resource relationship
	operators $\resAND$, $\resOR$, and $\resSEQ{t}$,
	where $\resSet$ is a set of medical resources required by the procedure $\procedure$.
\end{definition}

\begin{example}
	\label{ex:demand}
	As analyzed in Example~\ref{ex:relation}, the CT scan procedure requires three
	medical resources, i.e., CT machine, CT technician, and radiologist, with
	relationship given in formula~\eqref{eq:relation}.
	According to Definition~\ref{def:demand}, the resource demand $\resDemand$ of
	the CT scan procedure is	
	\begin{align}
	\label{eq:demand}
	\resDemand = \mathtt{CT\_machine} \ \resAND \ \mathtt{CT\_technician} \ \resSEQ{10} \ \mathtt{radiologist}.
	\end{align}	
\end{example}

We use the following two steps to automatically annotate medical resources
in executable medical guidelines:
\begin{enumerate}
	\item represent medical resource demands given by medical
	professionals with a map structure; and
	
	\item automatically annotate resource demands in \textit{states}
	and \textit{transitions} according to the resource demand map and raised medical
	procedures in corresponding \textit{states} and \textit{transitions}.
\end{enumerate}

We define the resource demand map structure as $(\mathtt{key}, \mathtt{value})$,
where the \textit{key} is medical procedures that are
represented by corresponding \textit{event} names in the medical guideline statecharts.
The \textit{value} of the resource demand map is the medical resource demand
of the corresponding \textit{key} (medical procedure).
We give the formal definition of the resource demand map structure in Definition~\ref{def:resMap}
and show the resource demand map of the simplified stroke scenario in Example~\ref{ex:resMap}.

\begin{definition}
	\label{def:resMap}
	Given an executable medical guideline model $\medicalGuideline$, a set
	of medical procedures $\procedureSet = \{ \procedure_1,
	\procedure_2, \dots, \procedure_n \}$ in the medical guideline $\medicalGuideline$,
	and a medical resource demand $\resDemand_i$ for each medical procedure $\procedure_i$,
	the medical resource demand map $\resMap$ is defined as
	\begin{align}
	\label{eq:resMap}
	\begin{split}		
	\resMap = &\{ (\procedure_1, \resDemand_1), \\		
	&(\procedure_2, \resDemand_2), \\		
	&\dots \dots  \\
	&(\procedure_n, \resDemand_n) \}.
	\end{split}
	\end{align}
\end{definition}

\begin{example}
	\label{ex:resMap}
	The simplified stroke statechart model shown in Fig.~\ref{fig:stroke}
	has two medical procedures $\mathtt{CTscan}$ and $\mathtt{givetPA}$.
	The medical resource demand of the $\mathtt{CTscan}$ procedure is given
	in formula~\eqref{eq:demand}. The $\mathtt{givetPA}$ procedure
	requires tPA fluid and a nurse.
	According to Definition~\ref{def:resMap}, the resource demand map of
	the simplified stroke scenario is	
	\begin{align}
	\label{eq:resMapStroke}
	\begin{split}		
	&\{ (\mathtt{CTscan}, \mathtt{CT\_machine}  \resAND  \mathtt{CT\_technician}  \resSEQ{10}  \mathtt{radiologist}), \\		
	&(\mathtt{givetPA}, \mathtt{tPA} \ \resAND \ \mathtt{nurse}) \}.
	\end{split}
	\end{align}	
\end{example}

The required demand represented in $\resMap$ is
independent of executable medical guideline models.
With the purpose of not affecting execution behaviors and validation/verification
results of medical verifiably correct executable medical guideline models, we annotate
medical resource demands by \yakindu\ statechart \textit{comments}. The annotation
is defined as follows.
\begin{definition}
	\label{def:annotation}
	Given a state $\state$ (or a transition $\transition$) in an executable medical
	guideline model $\medicalGuideline$,
	a set of medical procedures $\procedureSet_S = \{ \procedure_1,
	\procedure_2, \dots, \procedure_n \}$  modeled in state $\state$
	(or transition $\transition$), and	
	a medical resource demand map $\resMap$ of $\medicalGuideline$,
	the annotation of state $\state$ (or transition $\transition$) is represented as
	\begin{align}
	\label{eq:annotation}
	//@\mathtt{RES}: (\procedure_1, \resDemand_1), (\procedure_2, \resDemand_2),
	\dots, (\procedure_n, \resDemand_n).
	\end{align}	
\end{definition}

Based on the medical resource demand map and the
medical resource annotation definitions, we
annotate required medical resources in executable medical guideline statecharts with
following two steps:
first search each \textit{state} $\state$ (and transition $\transition$) in a given
medical guideline statechart $\medicalGuideline$;
second, if the actions of state $\state$ (or transition $\transition$) contain medical procedures
in the given medical resource demand map $\resMap$, add the annotation, i.e.,
formula~\eqref{eq:annotation}, to state $\state$ (or transition $\transition$).
Algorithm~\ref{alg:annotation} gives the details of the annotation procedure,
where the operation $\resMap' + (\procedure,\resDemand)$ in Line 6
inserts the element $(\procedure,\resDemand)$ into the map $\resMap'$.

\begin{algorithm}
	\caption{\textsc{Annotation}}
	\label{alg:annotation}
	\begin{algorithmic}[1]
		\REQUIRE An executable medical guideline model $\medicalGuideline$ and
		 a medical resource demand map $\resMap$ (formula~\eqref{eq:resMap}).
		\ENSURE The annotated medical guideline model $\medicalGuideline'$.
		
		\FOR{each state $\state$ or transition $\transition$ in $\medicalGuideline$}
			\STATE Define a map $\resMap'$
			\FOR{each raised action $\procedure$ in $\state$ or $\transition$}
				\STATE Find $\resDemand$ with key $\procedure$ in $\resMap$
				\IF{$\resDemand$ is not $\mathtt{NULL}$}
					\STATE $\resMap' = \resMap' + (\procedure,\resDemand)$
				\ENDIF			
			\ENDFOR
			\IF{$\resMap'$ is not empty}						
				\STATE Add an annotation in the format of formula~\eqref{eq:annotation} to state $\state$ or transition $\transition$
			\ENDIF
		\ENDFOR
		\RETURN $\medicalGuideline$
	\end{algorithmic}
\end{algorithm}

\begin{example}
	\label{ex:strokeAnnotation}
	Given the simplified stroke statechart model shown in Fig.~\ref{fig:stroke}
	and a resource demand map of formula~\eqref{eq:resMapStroke}.	
	The state $\mathtt{CT}$ has a medical procedure $\mathtt{CTscan}$.
	We use $\mathtt{CTscan}$ as the key to search the resource demand map given by
	formula~\eqref{eq:resMapStroke} and find resource demand
	$\mathtt{CT\_machine} \ \resAND \ \mathtt{CT\_technician} \ \resSEQ{10} \ \mathtt{radiologist}$.
	According to 	Definition~\ref{def:annotation}, we add the annotation
	``$//@\mathtt{RES}: (\mathtt{CTscan}, \mathtt{CT\_machine} \ \resAND \ \mathtt{CT\_technician} \ \resSEQ{10} \ \mathtt{radiologist})$''
	to state $\mathtt{CT}$.
	Similarly, we add the annotation ``$//@\mathtt{RES}: (\mathtt{givetPA}, \mathtt{tPA} \ \resAND \ \mathtt{nurse})$''
	to the transition from state $\mathtt{tPAcheck}$ to state $\mathtt{tPA}$.
	The annotated stroke statechart	model by Algorithm~\ref{alg:annotation}
	is depicted in Fig.~\ref{fig:strokeAnnotation}\footnote{\label{footnote:annotation}
	In \yakindu\ statecharts, we use ``AND'', ``OR'', and ``SEQ($t$)'' to replace
	$\resAND$, $\resOR$, and $\resSEQ{t}$ in the resource demand annotations, respectively.},
	where the annotated states and
	transitions are marked by red rectangles.
	
	\setlength{\intextsep}{5pt}
	\setlength{\columnsep}{0pt}
	\begin{figure}[ht]
		\centering
		\includegraphics[width = 0.49\textwidth]{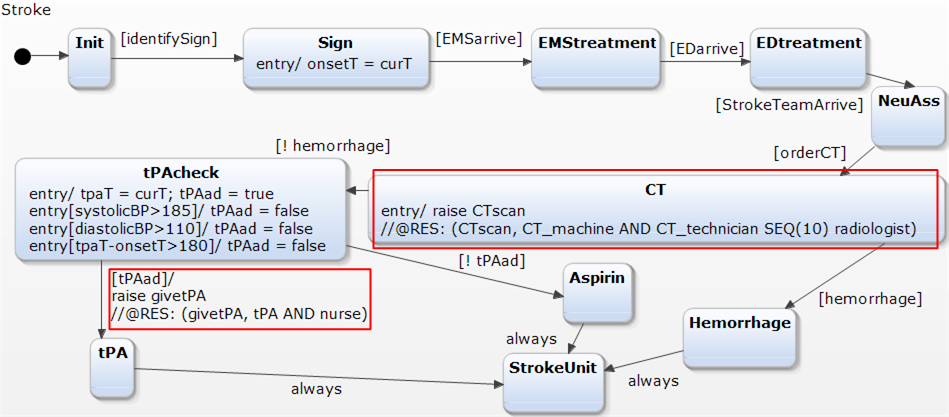}
		\caption[Annotated Statechart of Simplified Stroke Scenario]{Annotated Statechart of Simplified Stroke Scenario\footref{footnote:annotation}}
		\label{fig:strokeAnnotation}
	\end{figure}
\end{example}

%% file: resource.tex
For a given resource demand map $\resMap = \{ (\procedure, \resDemand) \}$ and its corresponding medical guideline models, 
we build resource demand statecharts in four steps:
(1) design a $\mathtt{Timer}$ statechart to record current system time;
(2) for each medical resource and each medical procedure in $\resMap$, declare a boolean
variable to denote whether a resource is available and
whether a procedure is able to execute, respectively;
(3) design a statechart for each basic resource relationship defined
in Section~\ref{subsec:relation};
and
(4) develop a procedure to automatically transform each element in
the resource demand map $\resMap$ to a statechart by composing the
basic resource relationship statecharts.

\begin{figure}[ht]
	\centering
	\includegraphics[width = 0.2\textwidth]{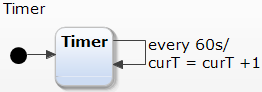}
	\caption{Timer Statechart}
	\label{fig:timer}
\end{figure}

For the $\mathtt{Timer}$ statechart,
we use an integer variable $\mathtt{curT}$ to denote current system time
and let a $\mathtt{Timer}$ statechart to increase current time $\mathtt{curT}$.
The $\mathtt{Timer}$ statechart only contains one state which has a self-loop
transition to increase current time $\mathtt{curT}$ by 1 every one time unit.
Fig.~\ref{fig:timer} depicts a $\mathtt{Timer}$ statechart with time
unit as minute, which increases $\mathtt{curT}$ by 1 $\mathtt{every \ 60s}$.

\setlength{\intextsep}{5pt}
\setlength{\columnsep}{0pt}

To represent resource demand related variables, we declare an interface
named $\mathtt{RES}$. For each unique resource $\res$ in resource demands
$\resDemand$ of a given resource map $\resMap$, we declare a boolean variable
$\var{\res}$ in the interface $\mathtt{RES}$ to denote whether the resource
$\res$ is available at current system time. 
The variable $\var{\res}$ has the same name as the corresponding resource $\res$
and the default value is $\mathtt{false}$ indicating the resource $\res$ is not
available initially.
The details on how to check whether a resource is available under given
resource schedules will be presented in Section~\ref{sec:availability}.

For each medical procedure $\procedure$ in the given resource map $\resMap$,
we declare a boolean variable $\var{\procedure}$ in the interface to denote
whether the procedure $\procedure$ is able to execute from resource demands
perspective. Similarly, the variable $\var{\procedure}$ has the same name
as the corresponding procedure $\procedure$ (\textit{event} name)
and the default value is $\mathtt{false}$ meaning that the procedure $\procedure$
can not be executed initially.
The procedure variables are used as anchors for integrating resource
demand statecharts with existing medical guideline models in Section~\ref{sec:integration}.
We use Example~\ref{ex:variable} to show the declared variable of the 
simplified stroke scenario.
\begin{example}
	\label{ex:variable}
	The medical resource demand map of the simplified stroke
	scenario is given in formula~\eqref{eq:resMapStroke}.
	The resource map contains two medical procedures, i.e., $\mathtt{CTscan}$ and $\mathtt{givetPA}$,
	and five medical resources, i.e., $\mathtt{CT\_machine}$, $\mathtt{CT\_technician}$,
	$\mathtt{radiologist}$, $\mathtt{tPA}$, and $\mathtt{nurse}$.
	The declared resource and procedure variables are shown in Fig.~\ref{fig:variable}.
	
	\setlength{\intextsep}{5pt}
	\setlength{\columnsep}{0pt}
	\begin{figure}[ht]
		\centering
		\includegraphics[width = 0.3\textwidth]{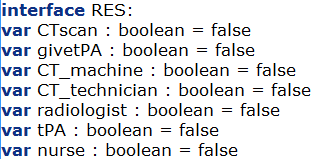}
		\caption{Resource and Procedure Variables}
		\label{fig:variable}
	\end{figure}	
\end{example}

According to Definition~\ref{def:demand}, resource demands are represented
as compositions of basic resource relationships.
To model resource demands with statecharts,
we first design a statechart to represent each basic resource relationship
defined in Section~\ref{subsec:relation}, then 
develop a procedure to automatically transform each resource demand
in $\resMap$ to a statechart by composing the basic resource relationship statecharts.

\begin{figure}[ht]
	\centering
	\includegraphics[width = 0.35\textwidth]{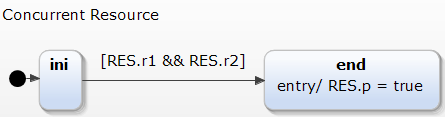}
	\caption{Concurrent Resource Statechart}
	\label{fig:andStatechart}
\end{figure}

Suppose a medical procedure $\procedure$ only requires two concurrent resources
$\res_1$ and $\res_2$, the resource demand map is
$\resMap = \{ (\procedure, \res_1 \ \resAND \ \res_2) \}$.
To model the concurrent resource demands, 
we build a statechart with two states $\mathtt{ini}$ and $\mathtt{end}$.
The statechart only contains one transition from state $\mathtt{ini}$ to state $\mathtt{end}$
with guard $\var{\res_1} \ \&\& \ \var{\res_2}$.
The entry action of the state $\mathtt{end}$ assigns the corresponding
procedure variable $\var{\procedure}$ to be $\mathtt{true}$.
The statechart ensures that the procedure $\procedure$ is able to
be executed only when
both $\res_1$ and $\res_2$ are available, i.e., $\res_1$ and $\res_2$ are
concurrent resources.
Fig.~\ref{fig:andStatechart} depicts the concurrent resource statechart.

\setlength{\intextsep}{5pt}
\setlength{\columnsep}{0pt}

\begin{figure}[ht]
	\centering
	\includegraphics[width = 0.35\textwidth]{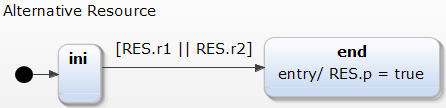}
	\caption{Alternative Resource Statechart}
	\label{fig:orStatechart}
\end{figure}

The alternative resource statechart is similar to the
concurrent resource statechart except that the transition guard
is $\var{\res_1} \ \| \ \var{\res_2}$, as shown in Fig.~\ref{fig:orStatechart}.
The alternative resource statechart ensures that the procedure $\procedure$ is able to
be executed when either one of $\res_1$ and $\res_2$ is available, i.e.,
$\res_1$ and $\res_2$ are alternative resources.

\setlength{\intextsep}{5pt}
\setlength{\columnsep}{0pt}

\begin{figure}[ht]
	\centering
	\includegraphics[width = 0.49\textwidth]{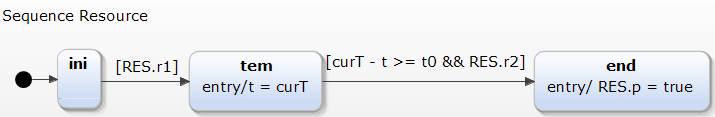}
	\caption{Sequence Resource Statechart}
	\label{fig:seqStatechart}
\end{figure}

Given a resource demand map $\resMap = \{ (\procedure, \res_1 \ \resSEQ{t_0} \ \res_2) \}$,
to model the sequence resource demands,
we build a statechart with three states $\mathtt{ini}$, $\mathtt{tem}$, and $\mathtt{end}$.
The statechart contains two transitions:
a transition from state $\mathtt{ini}$ to state $\mathtt{tem}$ with guard $\var{\res_1}$
and a transition from state $\mathtt{tem}$ to state $\mathtt{end}$ with
guard $\mathtt{curT-t>=t_0} \&\& \var{\res_2}$. In the transition guard,
$\mathtt{curT}$ is the current system time and $t$ is the time when the resource
$\res_1$ is available. The variable $t$ is assigned to the value of $\mathtt{curT}$
in the entry action of state $\mathtt{tem}$.
Similarly, the entry action of the state $\mathtt{end}$ assigns the corresponding
procedure variable $\var{\procedure}$ to be $\mathtt{true}$.
The statechart ensures that the procedure $\procedure$ is able to be executed
only when $\res_1$ is available and $\res_2$ is available after $t_0$ time units,
i.e., $\res_1$ and $\res_2$ are sequence resources.
Fig.~\ref{fig:seqStatechart} depicts the sequence resource statechart.

\setlength{\intextsep}{5pt}
\setlength{\columnsep}{0pt}

Given a resource demand map $\resMap$, we model each element
$(\procedure, \resDemand) \in \resMap$ with a statechart by
composing the three basic resource relationship statecharts shown in
Fig.~\ref{fig:andStatechart}, Fig.~\ref{fig:orStatechart}, and Fig.~\ref{fig:seqStatechart}
according to the resource demand $\resDemand$.
Example~\ref{ex:CTres} illustrates how to compose the basic
resource relationship statecharts for the CT scan scenario.

\begin{example}
\label{ex:CTres}
Given CT scan resource demand 
$\resDemand = \mathtt{CT\_machine} \ \resAND \ \mathtt{CT\_technician} \ \resSEQ{10} \ \mathtt{radiologist}$,
we apply the concurrent resource statechart shown in Fig.~\ref{fig:andStatechart}
to the demand $\mathtt{CT\_machine} \ \resAND \ \mathtt{CT\_technician}$\
and apply the sequence resource statechart  shown in Fig.~\ref{fig:seqStatechart}
to the rest part of $\resDemand$.
Hence, the statechart contains three states $\mathtt{ini}$, $\mathtt{tem}$, and $\mathtt{end}$.
The entry actions of state $\mathtt{tem}$ and state $\mathtt{end}$
are $t=\mathtt{curT}$ and $\mathtt{CTscan = true}$, respectively.
There are two transitions in the statechart:
a transition from state $\mathtt{ini}$ to state $\mathtt{tem}$ with guard $\mathtt{machine \&\& technician}$
and a transition from state $\mathtt{tem}$ to state $\mathtt{end}$ with
guard $\mathtt{curT-t>=10} \&\& \mathtt{radiologist}$.
The statechart resource model is shown in Fig.~\ref{fig:CTres}.

\setlength{\intextsep}{5pt}
\setlength{\columnsep}{0pt}
\begin{figure}[ht]
	\centering
	\includegraphics[width = 0.45\textwidth]{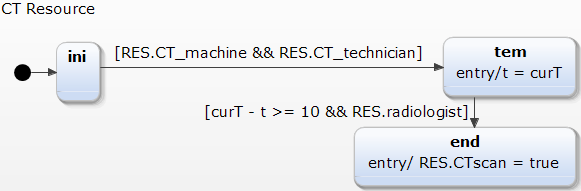}
	\caption{CT Resource Model}
	\label{fig:CTres}
\end{figure}
\end{example}

%% file: availability.tex
In this section, we first extract and convert medical resource available times from given resource available time schedules into a map structure, and then based on the map structure, we design a statechart and a Java class to model resource available times.

\subsection{Map Structure}

\begin{figure}[ht]
	\centering
	\includegraphics[width = 0.49\textwidth]{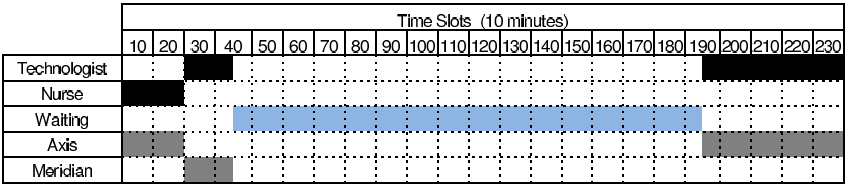}
	\caption{Schedule for MSC Bone Imaging Resources~\cite{Perez2011HSE}}
	\label{fig:schedule}
\end{figure}

In medical facilities, medical resource schedules are usually represented
as time tables. For instance, Fig.~\ref{fig:schedule} shows a schedule
of four medical resources, i.e., \textit{Technologist}, \textit{Nurse},
\textit{Axis}, and \textit{Meridian}, which are required for the 
MSC (mesenchymal stromal cell) bone imaging procedure~\cite{Perez2011HSE}.
Note that the \textit{Waiting} time is for patients that are not considered as medical resources in this paper.
In Fig.~\ref{fig:schedule}, each time slot represents 10 minutes and are
relative time. Medical resources are unavailable during the shaded time slots.
For example, the \textit{Technologist} is only available from time 0 to 20 minutes
and from 35 to 185 minutes.

\setlength{\intextsep}{5pt}
\setlength{\columnsep}{0pt}
 
To extract medical resource available times from given time tables,
we define the time interval and medical resource available time map as follows. 

\begin{definition}[Time Interval]
	\label{def:timeinterval}
	A time interval $\timeinterval$ is represented as $[\starttime, \endtime]$,
	where $\starttime$ and $\endtime$ are the start and end time points of
	the time interval $\timeinterval$, respectively.
\end{definition}

In the resource available time map structure $(\mathtt{key}, \overrightarrow{\mathtt{value}})$,
the \textit{key} is medical resources required by medical guidelines.
The \textit{value} of the resource available time map is time intervals
within which the corresponding resource (\textit{key}) is available.
As a medical resource may have multiple available time intervals,
we use an array of all available times intervals to
represent the \textit{value} in the resource available time map structure.
The formal definition of the resource available time map structure given below. 

\begin{definition}[Medical Resource Available Time Map]
	\label{def:avaMap}
	Given a set of medical resources $\resSet  = \{ \res_1, \res_2, \dots, \res_n\}$
	and a resource schedule $\schedule$,
	the medical resource available time model $\avaModel$ is defined as
	a map
	\begin{align}
	\label{eq:avaMap}
	\begin{split}		
	\avaMap = &\{ (\res_1, \{\timeinterval_1^1, \timeinterval_2^1, \dots, \timeinterval_{m_1}^1\}), \\		
	&(\res_2, \{\timeinterval_1^2, \timeinterval_2^2, \dots, \timeinterval_{m_2}^2\}), \\		
	&\dots \dots \dots \\
	&(\res_n, \{\timeinterval_1^n, \timeinterval_2^n, \dots, \timeinterval_{m_n}^n\}) \},
	\end{split}
	\end{align}
	where $\timeinterval_i^j$ is the $i$th time interval within which the medical resource
	$\res_j$ is available.
\end{definition}

The following example shows how to obtain the resource available time map from the simplified CT scan scenario.
\begin{example}
	\label{ex:avaMap}
	The CT scan scenario requires \textit{CT machines}, \textit{CT technicians},
	and \textit{radiologists}. Fig.~\ref{fig:CTschedule} shows an example
	schedule of CT scan resources in 40 minutes.
	Each time slot represents 5	minutes and are relative time.
	The shaded time slots are unavailable for corresponding resources.
	In Fig.~\ref{fig:CTschedule}, the \textit{CT Machine} has two available
	time intervals $[10,25]$ and $[35,40]$.
	Similarly, we can identify the available time intervals for the other two resources.
	The medical resource available time model of the CT scan example is
	\begin{align}
	\label{eq:avaMapEx}
	\begin{split}
	\map_\mathtt{CT} = \{
	&(\mathtt{CT\_machine}, \{[10,25], [35,40]\}), \\
	&(\mathtt{CT\_technician}, \{[0,10], [15,25], [35,40]\}), \\
	&(\mathtt{radiologist}, \{[0,15], [30,40]\})
	\}.
	\end{split}
	\end{align}
	
	\setlength{\intextsep}{5pt}
	\setlength{\columnsep}{0pt}
	\begin{figure}[ht]
		\centering
		\includegraphics[width = 0.49\textwidth]{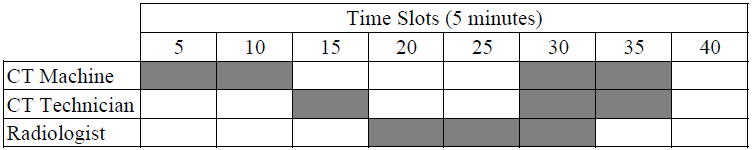}
		\caption{Schedule for CT Scan Resources}
		\label{fig:CTschedule}
	\end{figure}
\end{example}

\subsection{Statechart Model}

Given a medical resource available time map $\avaMap = \{ (\res, \{ \timeinterval \}) \}$, we develop a procedure
to represent $\avaMap$ with statecharts in three steps:
(1) use a configuration file to store resource available times specified in $\avaMap$;
(2) implement an external Java class $\mathtt{ResAva}$, which is supported by \yakindu\
statecharts, to read the resource available time configuration file and check whether
resources are available at current system time;
and
(3) design a statechart to access the Java class $\mathtt{ResAva}$ and represent
medical resource available times.
The approach has two advantages:
(1) when medical resource available times change which can happen quite often, we only need to update the resource available time configuration file without the need for statechart model changes;
(2) as there is only one statechart model for resource available times, the workload for validating and verifying the correctness of resource available time model is reduced.  

\begin{figure}[ht]
	\centering
	\includegraphics[width = 0.45\textwidth]{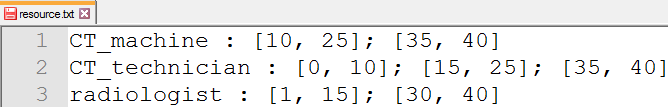}
	\caption{Resource Available Time Configuration File of CT Scan Scenario}
	\label{fig:config}
\end{figure}

In the resource available time configuration file, each element in the resource
available time map $\avaMap$ is stored in one line. For each map element,
we use colon ($:$) and semicolon ($;$) to separate the resource $\res$ and
its available time intervals $\{ \timeinterval \}$ and
each available time interval $\timeinterval$, respectively.
For example, the resource available time configuration file of the CT
scan scenario in formula~\eqref{eq:avaMapEx} is shown in Fig.~\ref{fig:config}.

\setlength{\intextsep}{5pt}
\setlength{\columnsep}{0pt}

\begin{figure}[ht]
	\centering
	\includegraphics[width = 0.49\textwidth]{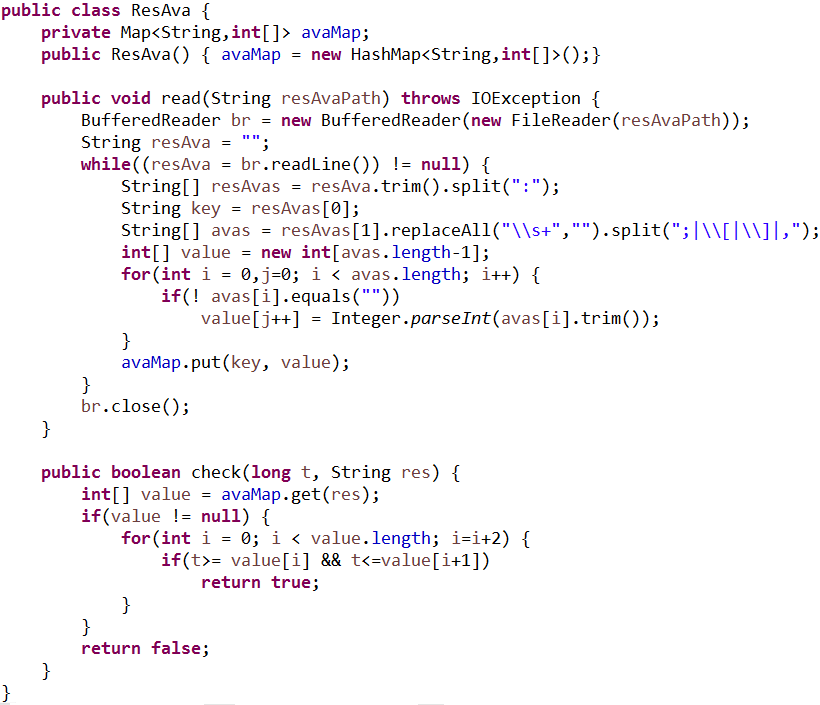}
	\caption{$\mathtt{ResAva}$ Java Class}
	\label{fig:code}
\end{figure}

\yakindu\ statechart supports external Java code, we implement
a Java class $\mathtt{ResAva}$ shown in Fig.~\ref{fig:code} to read medical resource available time configuration files and check resource availabilities.
The $\mathtt{ResAva}$ class provides two functions:
\begin{enumerate}
	\item $\mathtt{read(String \ resAvaPath)}$ reads the medical resource available 	time configuration file located in $\mathtt{resAvaPath}$;
	
	\item $\mathtt{check(long \ t, String \ res)}$ checks whether the medical resource
	$\mathtt{res}$ is available at time $t$.
\end{enumerate}

\setlength{\intextsep}{5pt}
\setlength{\columnsep}{0pt}

To model medical resource available times, we build a statechart with
only one state that has a self-loop transition with guard $\mathtt{true}$.
The transition ensures that resource available times are checked at each statechart
execution cycle.
The entry actions of the state call $\mathtt{read()}$ function to read
the configuration file and $\mathtt{check()}$ function for each resource
$\res$ in the given resource available time map $\avaMap$ to check resource
availabilities.
We use Example~\ref{ex:resource2} to show how to represent resource available time
for the CT scan scenario.

\begin{example}
	\label{ex:resource2}	
	Given the resource available time map $\avaMap$ of the CT scan scenario
	in formula~\eqref{eq:avaMapEx},
	the corresponding resource availability configuration file is shown in
	Fig.~\ref{fig:config}.
		
	To represent resource available times, we build the $\mathtt{Resource}$ statechart
	with only one state named $\mathtt{Res}$ which has a self-loop transition
	with guard $\mathtt{true}$.
	As the given resource available time map $\avaMap$ contains three medical resources
	$\mathtt{CT\_machine}$, $\mathtt{CT\_technician}$, and $\mathtt{radiologist}$,
	we add four	entry actions to the state $\mathtt{Res}$:
	\begin{enumerate}
		\item $\mathtt{ResAva.read(``/resource.txt"}$ reads the resource
		availability configuration file shown in Fig.~\ref{fig:config};
				
		\item $\mathtt{RES.CT\_machine = ResAva.check(curT, ``CT\_machine")}$ checks
		current availability of the resource $\mathtt{CT\_machine}$;
		
		\item {\small $\mathtt{RES.CT\_technician = ResAva.check(curT, ``CT\_technician")}$} checks
		current availability of the resource $\mathtt{CT\_technician}$;
		
		\item {\small $\mathtt{RES.radiologist = ResAva.check(curT, ``radiologist")}$} checks
		current availability of the resource $\mathtt{radiologist}$.		
	\end{enumerate}
	The resource statechart is shown in Fig.~\ref{fig:resoruce2}.	
	
	\setlength{\intextsep}{5pt}
	\setlength{\columnsep}{0pt}
	\begin{figure}[ht]
		\centering
		\includegraphics[width = 0.45\textwidth]{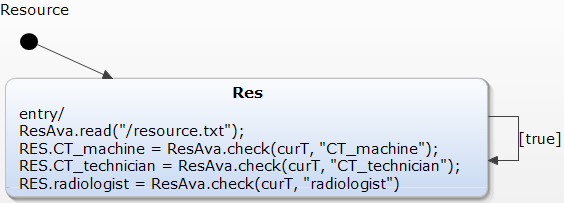}
		\caption{Resource Statechart of CT Scan Scenario}
		\label{fig:resoruce2}
	\end{figure}	
\end{example}

%% file: integration.tex
To clinically validate and formally verify the safety of medical guideline models
with consideration of medical resource available times and their relationships,
we need to integrate medical resource models with medical guideline statecharts.

According to the medical resource modeling approach presented
in Section~\ref{sec:resource}, for each medical procedure $\procedure$ in the given
resource demand map,
a boolean variable $\var{\procedure}$ is declared to denote
whether the procedure $\procedure$ is able to execute.
We use the declared procedure variable $\var{\res}$ as an anchor to
bridge the communication between
medical resource models and medical guideline statecharts
and modify medical guideline statecharts with following integration rules.
\begin{itemize}
	\item \textbf{Integration Rule 1}: For each transition $\transition$ with
	guard $\guard$, if it is annotated with
	``$//@\mathtt{RES}: (\procedure_1, \resDemand_1), (\procedure_2, \resDemand_2),
	\dots, (\procedure_n, \resDemand_n)$'',
	the guard $\guard$ is replaced with	
	$G = G \ \&\& \ \var{\procedure_1} \ \&\& \ \var{\procedure_2} \ \&\& \ \dots \ \&\& \ \var{\procedure_n}$;
	
	\item \textbf{Integration Rule 2}: For each state $\state$, if it is annotated
	with ``$//@\mathtt{RES}: (\procedure_1, \resDemand_1), (\procedure_2, \resDemand_2),
	\dots, (\procedure_n, \resDemand_n)$'',
	apply \textbf{Integration Rule 1} to all incoming transitions of the state $\state$
	with new guards.
\end{itemize}
We also design Algorithm~\ref{alg:integration} to automatically integrate
medical resource models with medical guideline statecharts.
Example~\ref{ex:integration} illustrates how we apply the integration rules to
integrate the resource demand given in formula~\eqref{eq:resMapStroke}
with the simplified stroke statechart.

\begin{algorithm}
	\caption{\textsc{Integration}}
	\label{alg:integration}
	\begin{algorithmic}[1]
		\REQUIRE An annotated medical guideline model $\medicalGuideline$.
		\ENSURE The integrated medical guideline model $\medicalGuideline'$.		
		
		\FOR{each state $\state$ in $\medicalGuideline$}
			\IF{$\state$ is annotated with ``$//@\mathtt{RES}: (\procedure_1, \resDemand_1), (\procedure_2, \resDemand_2), \dots, (\procedure_n, \resDemand_n)$''}
				\FOR{each incoming transition $\transition$ with guard $\guard$ of state $\state$}
					\STATE $G = G \ \&\& \ \var{\procedure_1} \ \&\& \ \var{\procedure_2} \ \&\& \ \dots \ \&\& \ \var{\procedure_n}$
				\ENDFOR
			\ENDIF
		\ENDFOR
				
		\FOR{each transition $\transition$ with guard $\guard$ in $\medicalGuideline$}
			\IF{$\transition$ is annotated with ``$//@\mathtt{RES}: (\procedure_1, \resDemand_1), (\procedure_2, \resDemand_2), \dots, (\procedure_n, \resDemand_n)$''}	
				\STATE $G = G \ \&\& \ \var{\procedure_1} \ \&\& \ \var{\procedure_2} \ \&\& \ \dots \ \&\& \ \var{\procedure_n}$	
			\ENDIF
		\ENDFOR		
		\RETURN $\medicalGuideline$
	\end{algorithmic}
\end{algorithm}

\begin{example}
\label{ex:integration}
	We integrate the resource models in Fig.~\ref{fig:CTres}
	with the annotated stroke statechart model in Fig.~\ref{fig:strokeAnnotation}.		
	The transition $\transition_1$ from state $\mathtt{tPAcheck}$
	to state $\mathtt{tPA}$ is annotated with ``$\mathtt{//@RES: (\mathtt{givetPA}, \mathtt{tPA} \ \resAND \ \mathtt{nurse})}$''
	and has guard $\guard_1 = \mathtt{tPAad}$. Based on \textbf{Integration Rule 1},
	the transition $\transition_1$' guard is set as
	$\guard_1 = \mathtt{tPAad} \ \&\& \ \mathtt{RES.givetPA}$.
	The state $\mathtt{CT}$ is annotated by
	``$\mathtt{//@RES:} \ (\mathtt{CTscan}, \mathtt{CT\_machine} \ \resAND \ \mathtt{CT\_technician} \ \resSEQ{10} \ \mathtt{radiologist})$''
	and only has one incoming transition $\transition_2$ with guard
	$\guard_2 = \mathtt{orderCT}$ from state $\mathtt{NeuAss}$.
	According to \textbf{Integration Rule 2}, we apply
	\textbf{Integration Rule 1} to the transition $\transition_2$ and
	set the guard as $\guard_2 = \mathtt{orderCT} \ \&\& \ \mathtt{RES.CTscan}$.
	Fig.~\ref{fig:strokeIntegration} shows the integrated stroke statechart,
	where the modified transitions are marked by red rectangles.
	
	\setlength{\intextsep}{5pt}
	\setlength{\columnsep}{0pt}
	\begin{figure}[ht]
		\centering
		\includegraphics[width = 0.49\textwidth]{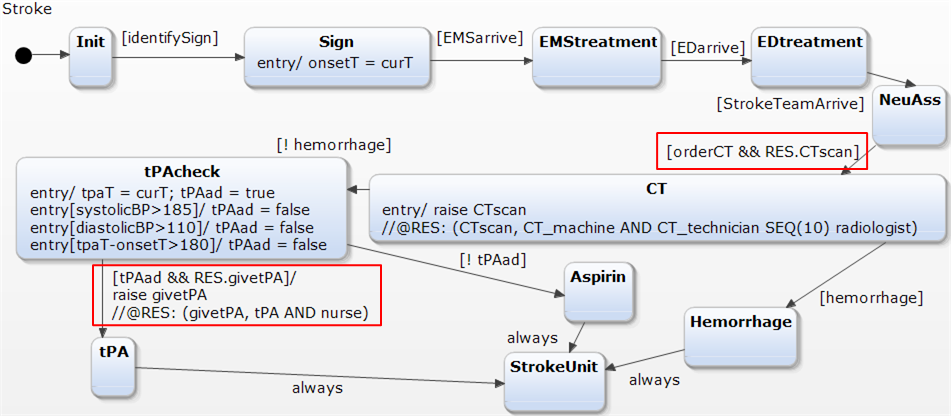}
		\caption{Integrated Statechart of Simplified Stroke Scenario}
		\label{fig:strokeIntegration}
	\end{figure}	
\end{example}

%% file: exp.tex
In this section, we use the stroke scenario to illustrate the effectiveness
and advantages of the proposed SMJV architecture which separately models medical resource available times and relationships and jointly verifies existing medical best practice guideline statecharts with resource models being integrated in.

\subsection{SMJV Architecture Validation}

The stroke statechart model given in Fig.~\ref{fig:stroke} has only focused on the CT scan and tPA administration procedures, but omitted the details of other medical procedures. To validate the effectiveness of the proposed architecture, we extend the simplified stroke model by considering following scenarios with different patient conditions: (1) if a patient's blood pressure is not within the range required by tPA administration, a blood pressure control procedure needs to be performed; (2) if tPA administration is approved within 3 hours from onset of stroke symptoms, an IV tPA procedure is performed; (3) if tPA administration is approved in the 3-6 hour window from the onset time, an IA tPA procedure is performed; and (4) if tPA is not approved, aspirin is given to patients.

\begin{figure}[ht]
	\centering
	\includegraphics[width = 0.49\textwidth]{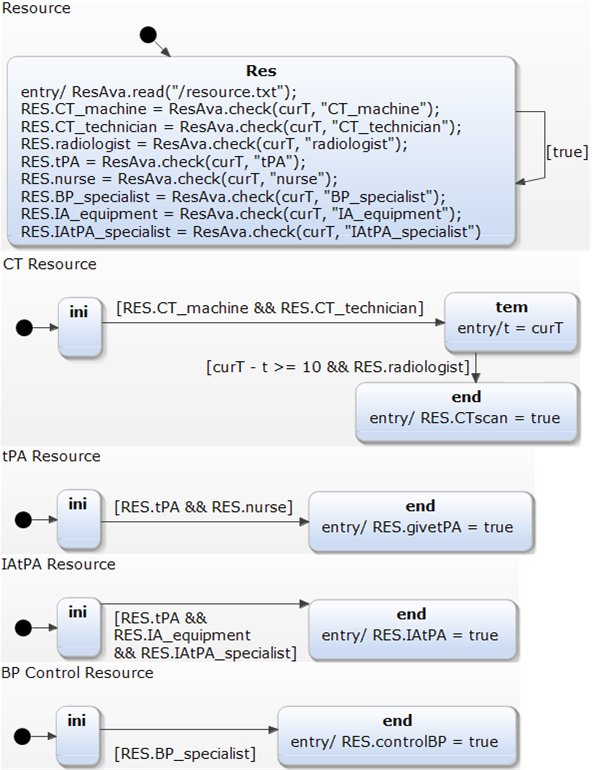}
	\caption{Stroke Resource Model}
	\label{fig:strokeFullRes}
\end{figure}

\begin{figure}[ht]
	\centering
	\includegraphics[width = 0.49\textwidth]{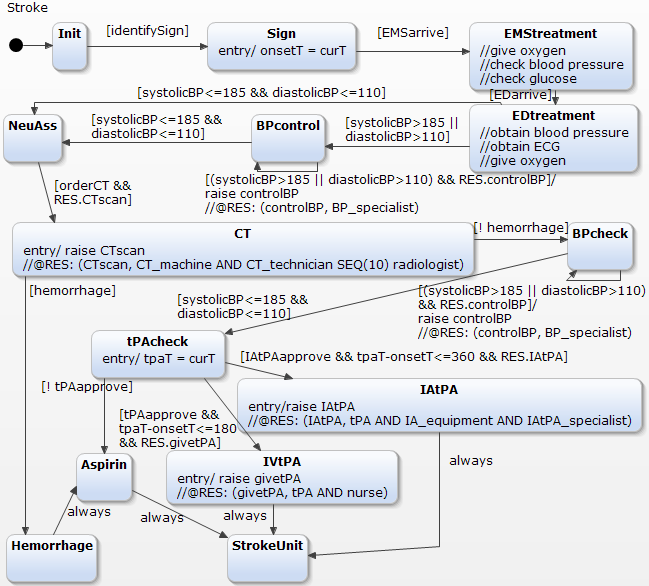}
	\caption{Integrated Stroke Statechart}
	\label{fig:strokeFull}
\end{figure}

We use the proposed approach to annotate resource demand, model resource
relationships and available times, and integrate resource models with
the extended stroke statechart model.
The resource models and integrated stroke statechart
are shown in Fig.~\ref{fig:strokeFullRes} and Fig.~\ref{fig:strokeFull}, respectively.

\setlength{\intextsep}{5pt}
\setlength{\columnsep}{0pt}

\setlength{\intextsep}{5pt}
\setlength{\columnsep}{0pt}

\begin{figure}[ht]
	\centering
	\includegraphics[width = 0.49\textwidth]{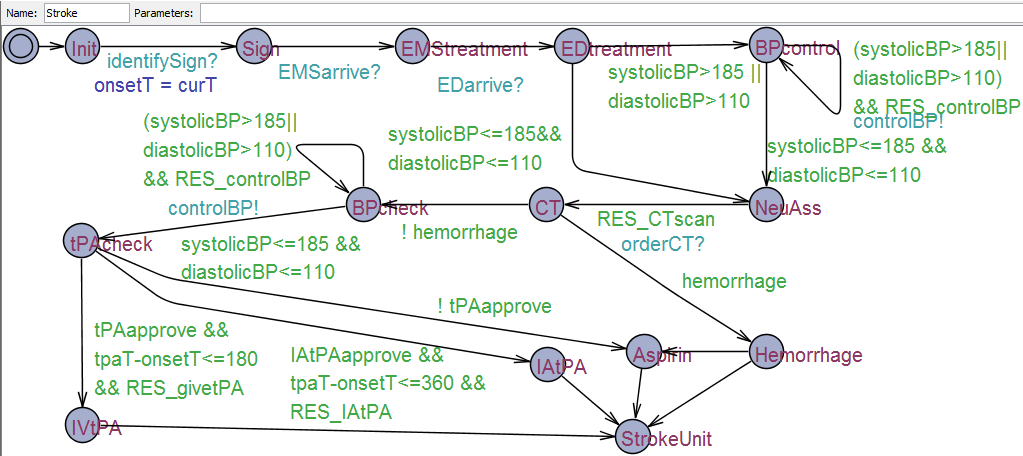}
	\caption{Stroke \uppaal\ Model}
	\label{fig:strokeFullU}
\end{figure}

To clinically validate and formally verify the safety of the stroke
statechart with the consideration of resource available times and relationships,
we run simulations of the integrated stroke statechart model (Fig.~\ref{fig:strokeFull})
through \yakindu, transform the integrated stroke model to an \uppaal\ model with
the \toolname\ tool~\cite{Guo2016ICCPS}
(Fig.~\ref{fig:strokeFullU}), and verify the safety properties in \uppaal.

\setlength{\intextsep}{5pt}
\setlength{\columnsep}{0pt}

In addition to the properties \textbf{P1} and \textbf{P2} given in
formula~\eqref{eq:P1} and formula~\eqref{eq:P2},
we also need to verify property \textbf{P3} that the IA tPA administration must
be completed within 6 hours from onset of stroke symptoms, i.e.,
\begin{align}
\label{eq:P3}
A[~] \ \mathtt{Stroke.IAtPA} \ imply \ \mathtt{tpaT}-\mathtt{onsetT}<=360
\end{align}

\begin{figure}[ht]
	\centering
	\includegraphics[width = 0.49\textwidth]{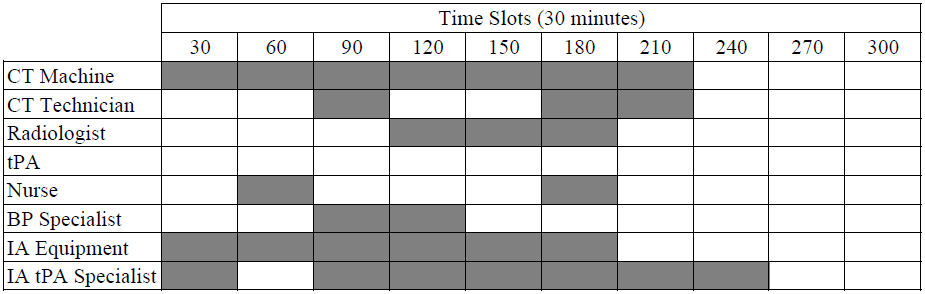}
	\caption{Stroke Resource Schedule}
	\label{fig:StrokeSchedule}
\end{figure}

Assume a patient's onset time of stroke symptom is 0,
the resource schedule is given in Fig.~\ref{fig:StrokeSchedule}.
Both simulation and verification results show that
the safety property \textbf{P1} and \textbf{P3} hold, but \textbf{P2} fails.

\setlength{\intextsep}{5pt}
\setlength{\columnsep}{0pt}

The case study demonstrates the following: (1) the proposed separately model and jointly verify approach are able to capture unsafety properties caused by temporarily unavailable resources in both clinical validation and formal verification processes; and (2) the approach can address patient care safety under different patient conditions by exploring different execution paths in medical guideline models with corresponding resource available times and their relationships being integrated in.

\subsection{Advantages of the SMJV Architecture}

The proposed SMJV architecture uses separate statecharts to model
medical resource available times and relationships and integrates
resources models with medical guideline statecharts to clinically
validate and formally verify safety properties in the presence of
resource demands and potential unexpected resource delays.
An alternative approach to address the medical resource demand issue
in existing medical guideline statecharts is to directly add
medical resource demands as transition guards or state constraints.
We call this method as direct modification approach.
Compared with the direct modification approach, the proposed
SMJV architecture has two major advantages.  First,  the separated resource demand statecharts and resource available time statecharts of common medical procedures can be reused by multiple medical guideline	models.  Second, the proposed approach requires minimal model modifications to accommodate frequent resource available time changes in emergency care environment.	 We use the CT scan scenario in the stroke care to show the two advantages.

Assume a resource available time schedule is given in Fig.~\ref{fig:StrokeSchedule}.
The direct modification approach models CT scan resource demands by
modifying the guard of transition from state $\mathtt{NeuAss}$ to state $\mathtt{CT}$ 
to be $\mathtt{orderCT \&\& CT\_machine>=210 \&\& CT\_techinician>=} \\ \mathtt{210\&\& radiologist>=180}$.
With the proposed approach, the CT scan resource demands are modeled with two
statecharts as depicted by Fig.~\ref{fig:CTres} and Fig.~\ref{fig:resoruce2} and
a resource available time configuration file shown in Fig.~\ref{fig:config}.  The resource models are integrated with the stroke guideline statecharts by
changing the guard of transition from state $\mathtt{NeuAss}$ to state $\mathtt{CT}$ from
$\mathtt{orderCT}$ to $\mathtt{orderCT \&\& CT\_scan}$.

On the surface, it seems that the SMJV requires more work than the direct modification approach.  However, once the resource demand model for CT scan is developed, it can be reused by other medical guideline models that require CT scan in their patient care, such as in the guideline model for acute appendicitis~\cite{AppendicitisGuideline}.  Even within the same medical guideline model, a medical procedure may be required at different patient care stages. For instance, in the stroke guideline model, there are two places that require the blood pressure control procedure as shown in Fig.~\ref{fig:strokeFull}, one when a 
patient arrives the emergency department, and the other before tPA is administrated.  But with the direct modification approach, for every medical guideline models where CT scan is needed, the corresponding transition guards has to be modified by adding the conjunction condition $\mathtt{orderCT \&\& CT\_machine>=210 \&\& CT\_techinician>=} \\ \mathtt{210\&\& radiologist>=180}$.

Furthermore, if the CT machine's available time changes from $[210, 300]$ to $[220,300]$, with the proposed approach, we only need to update the resource available time configuration file shown in Fig.~\ref{fig:config}. As none of the statechart models are changed, medical professionals only need to re-validate the resource available time configuration file and run our system to verify safety properties with updated resource available times. However, for the same available time change, the direct modification approach would need to modify all the transition guards leading to the CT scan. As the stroke statechart is modified to reflect updated resource availability, medical professionals are required to re-validate the stroke guideline model. For formal verification, the statechart needs to
be re-transformed to \uppaal\ timed automata to re-verify the safety properties.

%% file: conclusion.tex
The paper presents an approach to annotate resource demands in existing medical guideline models, separately model medical resource available times and relationships with statecharts,  and integrate resources models with executable medical guideline statecharts. The proposed architecture can be easily integrated into our existing platform~\cite{Guo2016ICCPS} and automatically transform medical guideline statecharts with resource demands to \uppaal\ timed automata so that medical guideline safety properties in the presence of temporarily
unavailable medical resources can be formally verified.

We have use a simplified stroke scenario as a case study to investigate 
the effectiveness and validity of the SMJV architecture.
The case study demonstrates: 
(1) the SMJV architecture allows different domain personals to make independent model modifications on resource available times as well as on medical guideline models, and requires minimal changes to integrate resource available time and relationship considerations into existing medical guideline models; (2) the SMJV architecture is able to identify unsafe properties caused by temporarily unavailable medical resources, and is easy adapt to frequent and unexpected resource available time changes; and (3) the separately modeled resource statecharts can be reused by multiple procedures within a medical guideline model and also by multiple medical guideline models.

%% file: main.bbl
\begin{thebibliography}{10}

\bibitem{alur1994theory}
R.~Alur and D.~L. Dill.
\newblock A theory of timed automata.
\newblock {\em Theoretical computer science}, 126(2):183--235, 1994.

\bibitem{IschemicStroke}
A.~S. Association.
\newblock Ischemic stroke.
\newblock
  \url{https://www.strokeassociation.org/idc/groups/stroke-public/@wcm/@hcm/documents/downloadable/ucm_309725.pdf}.

\bibitem{tpa-gold-standard}
A.~S. Association.
\newblock Stroke treatment.
\newblock
  \url{http://www.strokeassociation.org/STROKEORG/AboutStroke/Treatment/Stroke-Treatment_UCM_492017_SubHomePage.jsp}.

\bibitem{Balser2002Asbru}
M.~Balser, C.~Duelli, and W.~Reif.
\newblock Formal semantics of asbru – an overview.
\newblock {\em Proc. of the 6th Biennial World Conference on Integrated Design
  and Process Technology}, 5(5):1--8, 2002.

\bibitem{behrmann2004tutorial}
G.~Behrmann, A.~David, and K.~Larsen.
\newblock A tutorial on uppaal.
\newblock In {\em Formal Methods for the Design of Real-Time Systems}, pages
  200--236. Springer, 2004.

\bibitem{cayirli2003outpatient}
T.~Cayirli and E.~Veral.
\newblock Outpatient scheduling in health care: a review of literature.
\newblock {\em Production and operations management}, 12(4):519--549, 2003.

\bibitem{Christov2008Formally}
S.~Christov, B.~Chen, G.~S. Avrunin, L.~A. Clarke, L.~J. Osterweil, D.~Brown,
  L.~Cassells, and W.~Mertens.
\newblock Formally defining medical processes.
\newblock {\em Methods of Information in Medicine}, 47(5):392--398, 2008.

\bibitem{Christov2008}
S.~Christov, B.~Chen, G.~S. Avrunin, L.~A. Clarke, L.~J. Osterweil, D.~Brown,
  L.~Cassells, and W.~Mertens.
\newblock {\em Rigorously Defining and Analyzing Medical Processes: An
  Experience Report}, pages 118--131.
\newblock Springer Berlin Heidelberg, Berlin, Heidelberg, 2008.

\bibitem{conforti2008optimization}
D.~Conforti, F.~Guerriero, and R.~Guido.
\newblock Optimization models for radiotherapy patient scheduling.
\newblock {\em 4OR: A Quarterly Journal of Operations Research}, 6(3):263--278,
  2008.

\bibitem{AppendicitisGuideline}
S.~Di~Saverio, A.~Birindelli, M.~D. Kelly, F.~Catena, D.~G. Weber, M.~Sartelli,
  M.~Sugrue, M.~De~Moya, C.~A. Gomes, A.~Bhangu, F.~Agresta, E.~E. Moore,
  K.~Soreide, E.~Griffiths, S.~De~Castro, J.~Kashuk, Y.~Kluger, A.~Leppaniemi,
  L.~Ansaloni, M.~Andersson, F.~Coccolini, R.~Coimbra, K.~S. Gurusamy, F.~C.
  Campanile, W.~Biffl, O.~Chiara, F.~Moore, A.~B. Peitzman, G.~P. Fraga,
  D.~Costa, R.~V. Maier, S.~Rizoli, Z.~J. Balogh, C.~Bendinelli, R.~Cirocchi,
  V.~Tonini, A.~Piccinini, G.~Tugnoli, E.~Jovine, R.~Persiani, A.~Biondi,
  T.~Scalea, P.~Stahel, R.~Ivatury, G.~Velmahos, and R.~Andersson.
\newblock Wses jerusalem guidelines for diagnosis and treatment of acute
  appendicitis.
\newblock {\em World Journal of Emergency Surgery}, 11(1):34, Jul 2016.

\bibitem{fox1998disseminating}
J.~Fox, N.~Johns, and A.~Rahmanzadeh.
\newblock Disseminating medical knowledge: the proforma approach.
\newblock {\em Artificial Intelligence in Medicine}, 14(1-2):157 -- 182, 1998.

\bibitem{Guo2016ICCPS}
C.~Guo, S.~Ren, Y.~Jiang, P.-L. Wu, L.~Sha, and R.~Berlin.
\newblock Transforming medical best practice guidelines to executable and
  verifiable statechart models.
\newblock In {\em 2016 ACM/IEEE 7th International Conference on Cyber-Physical
  Systems (ICCPS)}, pages 1--10, April 2016.

\bibitem{gupta2008appointment}
D.~Gupta and B.~Denton.
\newblock Appointment scheduling in health care: Challenges and opportunities.
\newblock {\em IIE transactions}, 40(9):800--819, 2008.

\bibitem{resource_schecule}
R.~Hall and J.~Partyka.
\newblock Handbook of healthcare system scheduling.
\newblock Springer, 2012.

\bibitem{harel1987statecharts}
D.~Harel.
\newblock Statecharts: A visual formalism for complex systems.
\newblock {\em Science of computer programming}, 8(3):231--274, 1987.

\bibitem{yakindu}
Itemis.
\newblock Yakindu statechart tools (sct).
\newblock \url{https://www.itemis.com/en/yakindu/statechart-tools/}, July 2016.

\bibitem{StrokeGuideline}
E.~C. Jauch, B.~Cucchiara, O.~Adeoye, W.~Meurer, J.~Brice, Y.~Y.-F. Chan,
  N.~Gentile, and M.~F. Hazinski.
\newblock Part 11: adult stroke: 2010 american heart association guidelines for
  cardiopulmonary resuscitation and emergency cardiovascular care.
\newblock {\em Circulation}, 122(18 suppl 3):S818--S828, 2010.

\bibitem{Kim2010TII}
J.~Kim, I.~Kang, J.~Y. Choi, and I.~Lee.
\newblock Timed and resource-oriented statecharts for embedded software.
\newblock {\em IEEE Transactions on Industrial Informatics}, 6(4):568--578, Nov
  2010.

\bibitem{ss1}
M.~Liu and W.~Su.
\newblock An improvement of scheduling method for medical resource order and
  distribution.
\newblock In {\em 35th Chinese Control Conference (CCC)}, July 2016.

\bibitem{delayreport}
D.~I. McIsaac, K.~Abdulla, H.~Yang, S.~Sundaresan, P.~Doering, S.~G. Vaswani,
  K.~Thavorn, MPharm, and A.~J. Forster.
\newblock Association of delay of urgent or emergency surgery with mortality
  and use of health care resources: a propensity score–matched observational
  cohort study.
\newblock {\em Canadian Medical Association Journal}, 189(27), 2017.

\bibitem{Mckinley2011computer}
B.~A. McKinley, L.~J. Moore, J.~F. Sucher, S.~R. Todd, K.~L. Turner,
  A.~Valdivia, R.~M. Sailors, and F.~A. Moore.
\newblock Computer protocol facilitates evidence-based care of sepsis in the
  surgical intensive care unit.
\newblock {\em Journal of Trauma and Acute Care Surgery}, 70(5):1153--1167,
  2011.

\bibitem{goldenTime}
A.~C. of~Surgeons. Committee~on Trauma.
\newblock {\em ATLS advanced trauma life support program for doctors}.
\newblock American College of Surgeons, 2004.

\bibitem{patel1998representing}
V.~L. Patel, V.~G. Allen, J.~F. Arocha, and E.~H. Shortliffe.
\newblock Representing clinical guidelines in glif.
\newblock {\em Journal of the American Medical Informatics Association},
  5(5):467--483, 1998.

\bibitem{patrick2007improving}
J.~Patrick and M.~L. Puterman.
\newblock Improving resource utilization for diagnostic services through
  flexible inpatient scheduling: A method for improving resource utilization.
\newblock {\em Journal of the Operational Research Society}, 58(2):235--245,
  2007.

\bibitem{Perez2011HSE}
E.~Pérez, L.~Ntaimo, W.~E. Wilhelm, C.~Bailey, and P.~McCormack.
\newblock Patient and resource scheduling of multi-step medical procedures in
  nuclear medicine.
\newblock {\em IIE Transactions on Healthcare Systems Engineering},
  1(3):168--184, 2011.

\bibitem{Prince2013strokeIA}
E.~A. Prince, S.~H. Ahn, and G.~M. Soares.
\newblock Intra-arterial stroke management.
\newblock {\em Seminars in interventional radiology}, 30(03):282--287, 2013.

\bibitem{Terenziani2004GLARE}
P.~Terenziani, S.~Montani, A.~Bottrighi, M.~Torchio, G.~Molino, and
  G.~Correndo.
\newblock The glare approach to clinical guidelines: main features.
\newblock {\em Stud. Health Technol. Inform.}, pages 162--166, 2004.

\bibitem{Tu2001EON}
S.~W. Tu and M.~A. Musen.
\newblock Modeling data and knowledge in the eon guideline architecture.
\newblock {\em Medinfo}, pages 280--284, 2001.

\bibitem{vermeulen2009adaptive}
I.~B. Vermeulen, S.~M. Bohte, S.~G. Elkhuizen, H.~Lameris, P.~J. Bakker, and
  H.~La~Poutr{\'e}.
\newblock Adaptive resource allocation for efficient patient scheduling.
\newblock {\em Artificial intelligence in medicine}, 46(1):67--80, 2009.

\bibitem{WuTreatment2014}
P.~Wu, D.~Raguraman, L.~Sha, R.~Berlin, and J.~Goldman.
\newblock A treatment validation protocol for cyber-physical-human medical
  systems.
\newblock In {\em Software Engineering and Advanced Applications (SEAA), 2014
  40th EUROMICRO Conference on}, pages 183--190, Aug 2014.

\bibitem{WuWorkflow2015}
P.~Wu, L.~Sha, R.~B. Berlin, and J.~M. Goldman.
\newblock Safe workflow adaptation and validation protocol for medical
  cyber-physical systems.
\newblock In {\em 2015 41st Euromicro Conference on Software Engineering and
  Advanced Applications}, pages 464--471, Aug 2015.

\end{thebibliography}
